\documentclass[aps,prl,twocolumn,showpacs]{revtex4}
\usepackage[T1]{fontenc}
\usepackage[latin1]{inputenc}
\usepackage{graphicx}
\usepackage{hyperref}
\usepackage{amsmath}



\begin{document}

\title{$q$-Breathers in finite lattices: nonlinearity and weak disorder}
\author{M.~V.~Ivanchenko}

\affiliation{Department of Applied Mathematics, University of Leeds, LS2 9JT, Leeds, United Kingdom}

\begin{abstract}
Nonlinearity and disorder are the recognized ingredients of the lattice vibrational dynamics, the factors that could be diminished, but never excluded. We generalize the concept of $q$-breathers -- periodic orbits in nonlinear lattices, exponentially localized in the reciprocal linear mode space -- to the case of weak disorder, taking the Fermi-Pasta-Ulan chain as an example. We show, that these nonlinear vibrational modes remain exponentially localized near the central mode and stable, provided the disorder is sufficiently small. The instability threshold depends sensitively on a particular realization of disorder and can be modified by specifically designed impurities. Basing on it, an approach to controlling the energy flow between the modes is proposed. The relevance to other model lattices and experimental miniature arrays is discussed.
\end{abstract}


\maketitle
Nonlinearity and disorder are ubiquitous and unavoidable features of discrete extended systems, the key players in a wealth of fundamental dynamical and statistical physical phenomena such as thermalization, thermal conductivity, wave propagation, electron and phonon scattering. Studying lattice vibrational modes is vital to gain full understanding of these problems. Nonlinearity induces interaction between normal modes and energy sharing if strong enough (known as the Fermi-Pasta-Ulam (FPU) problem) \cite{fpu}, and time-periodic exponential localizations in direct space (discrete breathers) \cite{flach_review}. Linear systems with disorder support another generic class of exponentially localized modes (Anderson modes) \cite{anderson}. At the same time a satisfactory full understanding of the {\it concurrent effect} of nonlinearity and disorder is missing. This gap is being progressively filled for {\it strongly disordered and weakly nonlinear} lattices by intensive research on continuation of Anderson modes into nonlinear regime \cite{anderson_continuation}, wavepacket spreading \cite{spreading}, light propagation in photonic lattices \cite{segev}, and Bose-Einstein condensate (BEC) localization in random optical potentials \cite{lahini}. 

Little, however, is known on how the systems with {\it pronounced nonlinearity and weak disorder} behave. Remarkably, it is a demand in a number of experimental and applicational contexts, beside a challenge from theory. Micro- and nano-electro-mechanical systems are rapidly developing components in microinstruments design (ultrafast sensors, radio frequency filters, added mass sensors) \cite{Roukes1}. Their array structures offer broadband excitations, elastic waves, and effects of dispersion to be utilized \cite{Sievers,Roukes2}. Importantly, they are often suggested to operate in the nonlinear regime, while maturing technology reduces fabrication errors, hence diminishing spatial disorder. On the atomic scale, three-dimensional gold nano-cluster structures are found to be active and selective catalysts for a variety of chemical reactions, the surface vibrational modes being possible initiators \cite{gold_clusters}. 
%Light propagation and cold atoms dynamics in random optical media \cite{segev,lahini} add to the list of motivating physical problems.  

One of the fundamental types of nonlinear oscillatory modes is q-breathers (QBs) - exact time-periodic solutions continued from linear modes and preserving exponential localization in the linear normal mode coordinates. Originally proposed to explain the FPU paradox (the energy locking in low-frequency modes of a weakly nonlinear chain, recurrencies, and size-dependent stochasticity thresholds) \cite{we_qb}, they have been discovered in two and three dimensional acoustic lattices, scaled to infinite systems, found in discrete nonlinear Schr\"{o}dinger (DNLS) arrays, and quantum QBs were observed in the Bose-Hubbard chain \cite{qqb}. Recently, QBs have been suggested as major actors in a BEC pulsating instability and a four-wave mixing process in a nonlinear crystal \cite{qb_in_crystal}. 
%The present principal challenge is whether QBs survive disorder and adequately describe vibrational dynamics of a lattice, meeting this %level of physical plausibility.

In this paper we extend the concept of QBs to random nonlinear media, exemplifying in the FPU chain. The cornerstones of our approach are continuation of QBs into non-zero 'frozen' disorder, taking a nonlinear localized solution as a seed, and statistical analysis of continued solutions. We show that QBs demonstrate the crossover from the exponential localization near the central mode to plateaus at a distance. The average stability threshold in nonlinearity remains the same in the first order of approximation. In contrast, the standard deviation increases linearly with disorder, manifesting a high sensitivity on realizations. We analyze the effect of the harmonic in space inhomogeneities and discuss the energy flow control by impurities design. 

The FPU-$\beta$ chain of $N$ equal masses, coupled by
springs with disorder in linear coefficients and quartic nonlinearity in potential, is described by the Hamiltonian
\begin{equation}
\label{eq1}
\begin{aligned}
& H=\frac{1}{2}\sum\limits_{n=1}^{N} p_{n}^2
+\sum\limits_{n=1}^{N+1}\left[\frac{1}{2}(1+D\kappa_{n})(x_{n}-x_{n-1})^2\right.\\
& \left.+\frac{\beta}{4}(x_{n}-x_{n-1})^4 \right] 
\end{aligned}
\end{equation}
where $x_{n}(t)$ is the displacement of the
$n$-th particle from its original
position, $p_{n}(t)$ its momentum, $x_{0}=x_{N+1}=0$,
$\kappa_{n}\in[-1/2,1/2]$ are random, uniformly distributed, and uncorrelated with $\left\langle \kappa_{n}\kappa_{m}\right\rangle=\sigma^2_{\kappa}\delta_{n,m}$, $\sigma^2_\kappa=1/12$ in our case. 
 A canonical transformation
$x_{n}(t)=\sqrt{\frac{2}{N+1}}\sum\limits_{q=1}^N Q_{q}(t)
\sin{\left(\frac{\pi q n}{N+1}\right)}$
 defines the
reciprocal wave number space with $N$ normal mode coordinates
$Q_{q}(t)$, being solutions to the linear disorder-free case. 
The normal mode space is spanned by $q$ and represents
a chain similar to the situation in real space.
The equations of motion read
\begin{equation}
\label{eq2}
\begin{aligned}
 &\Ddot{Q}_{q}+\omega_{q}^2
 Q_{q}=-\frac{\beta}{2(N+1)}\sum\limits_{p,r,s}^N
 C_{q,p,r,s}\omega_q\omega_p\omega_r\omega_s\\ 
 & \times Q_{p}Q_{r}Q_{s}-\frac{D}{\sqrt{N+1}}\sum\limits_p^N \omega_q\omega_p K_{q,p} Q_p\;.
 \end{aligned}
\end{equation}
Here $\omega_{q}=2\sin{\frac{\pi q}{2(N+1)}}$ are the normal mode frequencies. 
The coupling coefficients
$C_{q,p,r,s}$ \cite{we_qb}
induce the selective nonlinear interaction between distant
modes and $K_{q,p}=\frac{2}{\sqrt{N+1}}\sum\limits_{n=1}^{N+1}\kappa_n
\cos\frac{\pi q(n-1/2)}{N+1}\cos\frac{\pi p(n-1/2)}{N+1}$ reflect the all-to-all 
linear interaction due to disorder.

Our methodology consists of two steps. Firstly, we take a known QB solution for non-zero nonlinearity \cite{we_qb}. A particular realization of $\{\kappa_n\}$ is chosen and $d=D/\sqrt{N+1}$ regarded as the disorder parameter. Together with the nonlinearity parameter $\nu=\beta/(N+1)$, it is assumed to be small: $\nu, d\ll 1$. Then, an asymptotic expansion in powers of $\{\nu, d\}$ is developed. Linear stability analysis is based on the constructed solution. Secondly, we address the statistical properties of the QB solution and instability threshold calculating respective averages and variances.

Continuation of QBs to $\beta,D\neq 0$ from $\beta\neq0, D=0$ employs the same technique as to $\beta\neq 0,D=0$ from $\beta=D=0$ \cite{we_qb}. 
For $\nu, d<<1$ and small amplitude excitations the $q$-oscillators get effectively decoupled, their
harmonic energy $E_{q}=\frac{1}{2}\left(\dot{Q}_{q}^2+\omega_{q}^2
Q_{q}^2\right)$ being almost conserved in time. Single $q$-oscillator excitations
$E_{q} \neq 0$ for $q\equiv q_0$ only are trivial time-periodic
and $q$-localized solutions for $\beta=D=0$.

For the disorder-free chain such periodic orbits can
be continued into the nonlinear case at fixed total energy \cite{we_qb}
because the non-resonance condition
$n\omega_{q_0} \neq \omega_{q \neq q_0}$ ($n$ being an integer) holds for any finite 
size \cite{Tiziano} and the Lyapunov theorem \cite{lyapunov} applies.
Same ideas are expected to work for $d\ll 0$, as the spectrum remains non-resonant with the probability $1$ \cite{anderson_continuation}. 
Such continuation succeeded for all parameters we took.

\begin{figure}[t]
{\centering
\resizebox*{0.95\columnwidth}{!}{\includegraphics{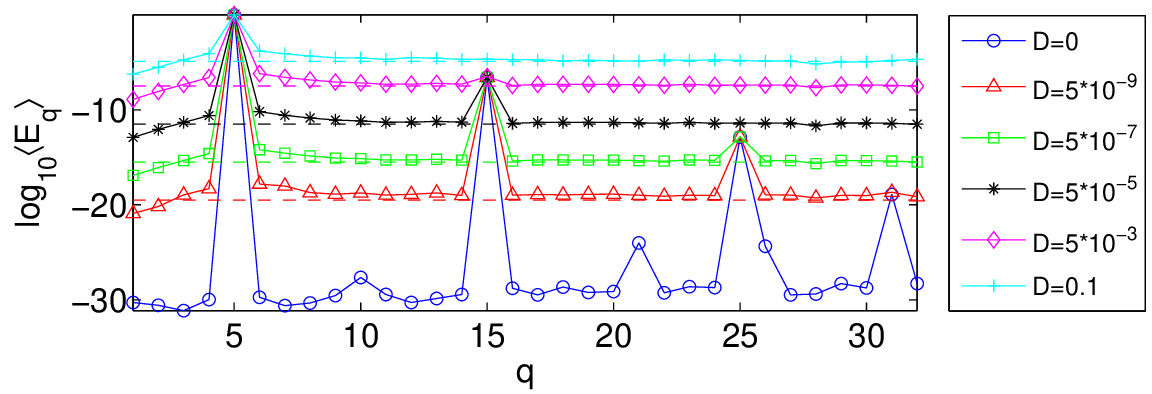}}}
{\caption{The average mode energy distribution in QBs with $q_0=5, \beta=0.01, N=32$ under increasing disorder. Dashed lines are theoretical estimates (\ref{eq2c}).}\label{fig1}}
\end{figure}

We continue QBs from nonlinear disorder-free solutions increasing $D$ and keeping a particular random realization $\{\kappa_n\}$ fixed. 
The total energy of the chain is $E=1$ in all examples, and $100$ realizations of disorder are taken and used for averaging. Dependence of the average QB energy distribution on the level of disorder is reported in Fig.\ref{fig1}. We observe the characteristic QB exponentially localized profile on the almost flat disorder-induced background. The height of the plateau grows with $D$, gradually absorbing localized modes. For $\beta=0.01, q_0=5,$ and $N=32$ the second large mode is overcome near $D=0.01$, and interpolation predicts the central one remains well above the background even for $D>2$, when some linear elasticity coefficients may become negative with a non-zero probability. 

Let us recall, that in case $D=0$ the QB solution $\hat{Q}_{q}^{NL}(t)$ with 
a low-frequency seed mode number $q_0$ can be written as an asymptotic expansion in
powers of the small nonlinearity parameter $\nu$ \cite{we_qb}. The energies of the
modes $q_0$, $3q_0$,\dots,$(2n+1)q_0$,\dots$\ll N$
read
\begin{equation}
\label{eq4}
E_{(2n+1)q_0}^{NL}=\lambda^{2n}E_{q_0}\;,\;
\lambda=\frac{3\beta
E_{q_0}(N+1)}{8\pi^2 q_0^2}\;\;,
\end{equation}
and the frequency $\omega^{NL}=\omega_{q_0}(1+9/4 \nu E_{q_0})$.
Now we develop a perturbation theory to (\ref{eq2}) in terms of the small disorder parameter $d$: $\hat{Q}_q(t)={Q}_q^{(0)}(t)+d{Q}_q^{(1)}(t)+\ldots$, its frequency being $\hat{\omega}=\omega^{(0)}+d\omega^{(1)}+\ldots$, substituting ${Q}_q^{(0)}(t)=\hat{Q}_{q}^{NL}(t)$ and $\omega^{(0)}=\omega^{NL}$. In the first order approximation (\ref{eq2}) becomes the equation of a forced oscillator:
$\Ddot{Q}_{q}^{(1)}+\omega_{q}^2 Q_{q}^{(1)}=-\omega_q\omega_{q_0} K_{q,q_0} Q_{q_0}^{(0)}$. It follows that all modes get excited by disorder, their amplitude the bigger the closer its frequency to $\omega_{q_0}$:
\begin{equation}
\label{eq2b}
\begin{aligned}
 A_{q}^{(1)}=-\frac{\omega_q\omega_{q_0}}{\omega_q^2-\omega_{q_0}^2} K_{q,q_0} A_{q_0}, \ q\neq q_0,
 \end{aligned}
\end{equation}
the frequency being $\hat{\omega}=\omega_{q_0}(1+9/4\nu E_{q_0}+d/2K_{q,q_0})$. As $\left\langle K_{q,q_0}\right\rangle=0$, the first order corrections in $d$ vanish for the averages $\left\langle A_{q}^{(1)}\right\rangle=\left\langle \omega^{(1)}\right\rangle=0$. Naturally, the variances are non-zero as the amplitude and frequency corrections vary depending on realization of $\{\kappa_n\}$. The mode energy (averaged over period of a QB) approximately separates into nonlinearity and disorder-induced parts $E_q\approx E_q^{NL}+E_q^{DO}$, where

\begin{equation}
\label{eq2c}
\begin{aligned}
\left\langle E_{q}^{DO}\right\rangle=\frac{d^2 E_{q_0}\omega_q^4}{2(\omega_q^2-\omega_{q_0}^2)^2}\left\langle K_{q,q_0}^2\right\rangle=\frac{d^2 E_{q_0} \sigma_\kappa^2\omega_q^4}{2(\omega_q^2-\omega_{q_0}^2)^2}
 \end{aligned}
\end{equation}

\begin{figure}[t]
{\centering
\resizebox*{0.95\columnwidth}{!}{\includegraphics{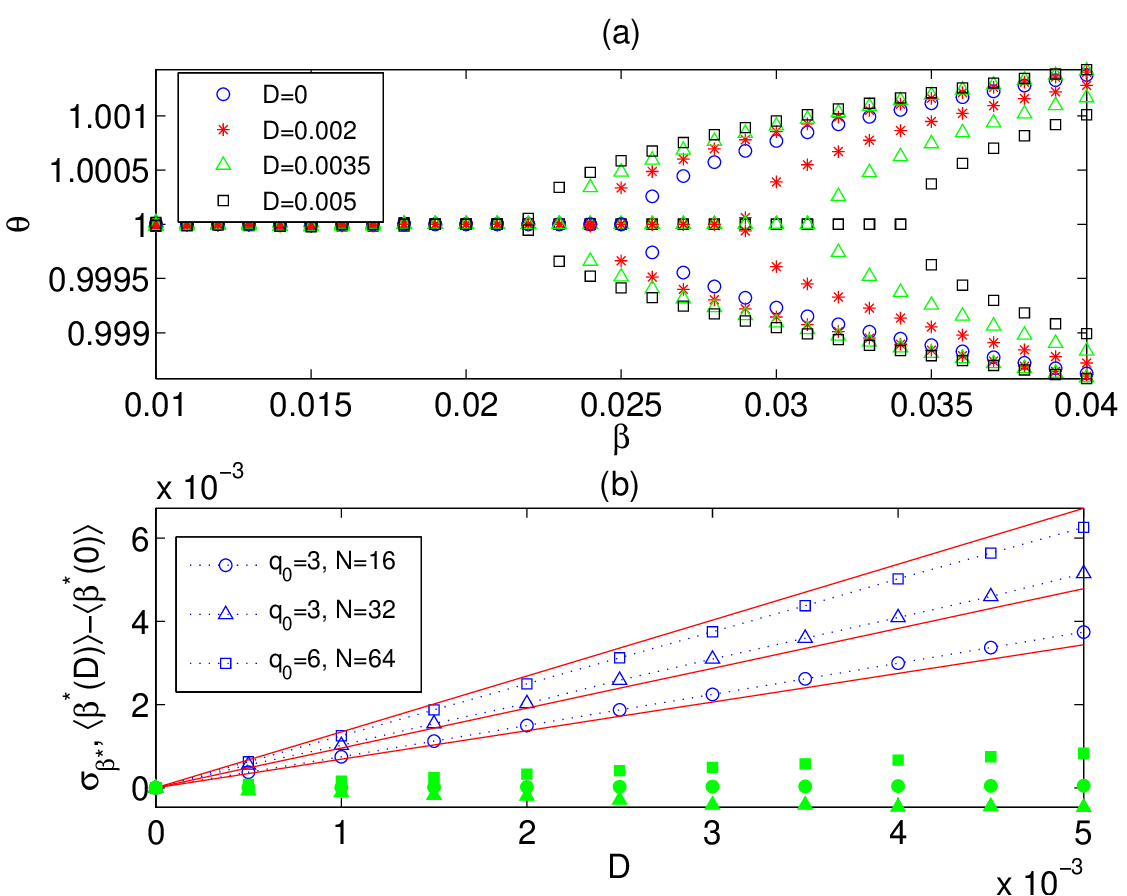}}}
{\caption{(a) The maximal and minimal absolute values of the eigenvalues of QBs with $q_0=6, N=64$ for two realizations of $\{\kappa_n\}$. For one realization the instability threshold in nonlinearity is increasing with $D$, for another -- decreasing. (b) Empty markers, dotted line: dependence of the variance of the instability threshold $\sigma_{\beta^*}$ on the disorder strength. Solid lines are theoretical estimates (\ref{eq5}). Filled markers: $\langle\beta^*(D)-\langle\beta^*(0)\rangle$.} 
\label{fig2}}
\end{figure}

Two limit cases are of particular interest: (i) $q \gg q_0$, then $\left\langle E_q^{DO}\right\rangle\approx d^2E_{q_0}\sigma_\kappa^2/2$ that gives a $q$-independent plateau energy (dashed lines in Fig.\ref{fig1}), and (ii) $q=q_0+1$, then  $\left\langle E_q^{DO}\right\rangle\approx d^2\sigma_\kappa^2\omega^2_{q_0}(N+1)^2E_{q_0}/2$, that yields the QB localization criterion $E_{q_0}\gg E_{q_0+1}$ if $\omega_{q_0}D\sigma_\kappa\ll 2 E_{q_0}/(N+1)$. Expressions (i) and (\ref{eq4}) predict the crossover between the exponential localization and the plateau at $q_c\approx\left(\ln{\frac{D\sigma_\kappa^2}{2(N+1)}}/\ln{\lambda}+1\right)q_0$. Expression (ii) suggests the 'small' $D\sigma_\kappa\ll \sqrt{2\pi^3/(N+1)^3}$ and 'large' $D\sigma_\kappa\gg \sqrt{8(N+1)}$ disorder criteria, the former implying that all QBs are localized and the latter that even QBs with $q_0=1$ are delocalized. It also recovers the boundary $q_0^*\propto\sqrt{N+1}$ between the localized (QBs) and delocalized in the q-space (but localized in the direct space) solutions (Anderson modes), that agrees with the previous results \cite{ishii}. Note, that the parameters taken in this paper correspond to the 'small' disorder.

The linear stability of the continued periodic orbits 
is determined by linearizing the phase space flow 
around them and computing the eigenvalues $\theta_i, i=\overline{1,2N}$
of the corresponding symplectic Floquet matrix \cite{we_qb}.
A QB is stable if $\left|\theta_i\right|=1,\ \forall i$.
The maximal and minimal absolute values of $\theta_i$ of QBs with $q_0=6, N=64$ for several increasing values of $D$ and two different $\{\kappa_n\}$ are plotted vs. $\beta$ in Fig.\ref{fig2}(a). Remarkably, while the instability threshold varies monotonously with $D$, it may not only decrease, but increase as well, depending on a particular $\{\kappa_n\}$. Moreover, stabilizing realizations are common, neatly balancing destabilizing ones. The observed deviation of the average instability threshold $\left\langle \beta^*\right\rangle$ from the disorder-free value $\beta_0^*$ was much smaller that the variance (Fig.\ref{fig2}(b)). The latter grows almost linearly in $D$,  up to $\sigma_{\beta^*}\approx0.25\beta_0^*$, as seen for $q_0=6, N=64$ (Fig.\ref{fig2}(b); note that for larger $D$ the linear fit may become violated, due to the lower bound $\beta^*>0$). 

%\begin{figure}[t]
%{\centering
%\resizebox*{0.90\columnwidth}{!}{\includegraphics{figs/fig2.eps}}}
%{\caption{(Color online) The variance of the instability threshold $\sigma_{\beta^*}$ vs. $D$ for QBs with $q_0=3, N=16$ (circles), $q_0=3, N=32$ %(triangles), and $q=6, N=64$ (squares). Solid lines are theoretical estimates (\ref{eq5}).} \label{fig4}}
%\end{figure}

\begin{figure}[t]
{\centering
\resizebox*{0.95\columnwidth}{!}{\includegraphics{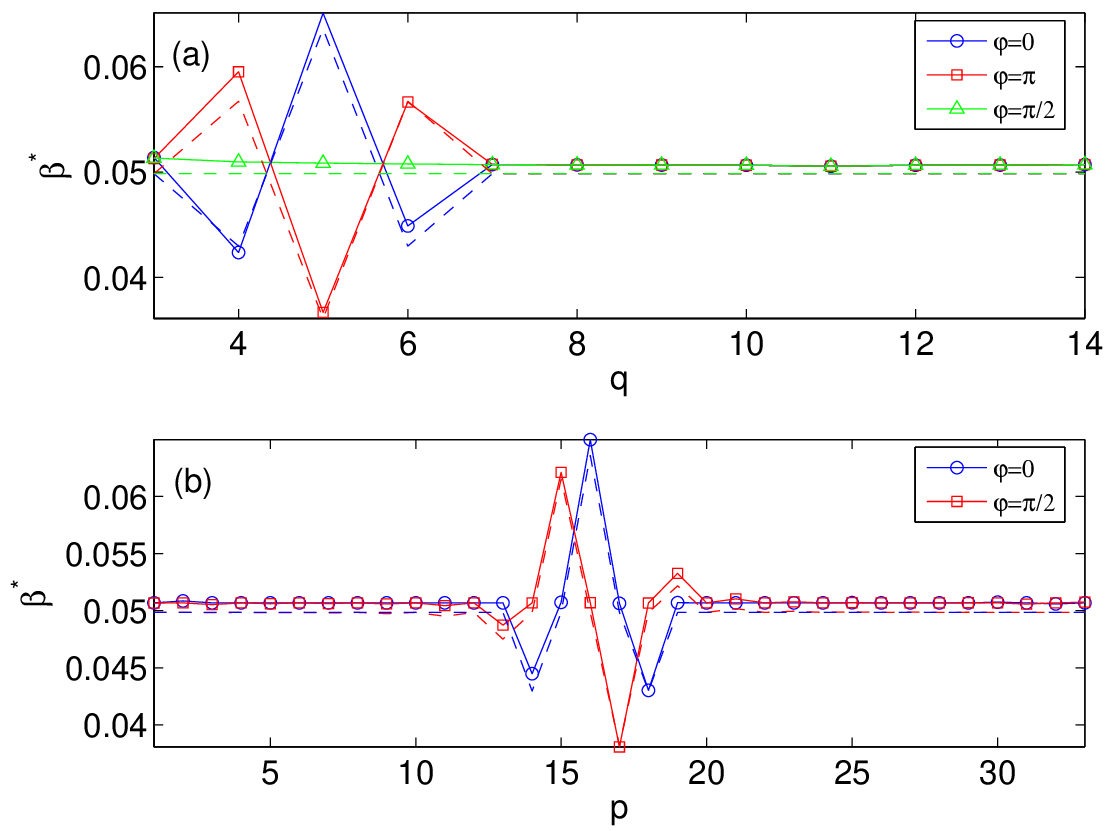}}}
{\caption{QB stability in case of spatially harmonic modulations:
$N=32, D=0.0025$ and (a) $p=10$, the central mode $q$ is changed, (b) the central mode $q=8$, the modulation wave number $p$ is varied. Dashed lines are theoretical estimates.}\label{fig5}}
\end{figure}

The monotonous dependence of the instability threshold on $D$ suggests that it is caused by the same resonance with the modes $q_0\pm1$ as in the disorder-free case. Let us explore the impact of disorder on this bifurcation. Linearizing equations of motion (\ref{eq2}) around a QB solution $Q_{q}=\hat{Q}_{q}(t)+\xi_{q}(t)$, one gets
 \begin{equation}
  \label{eq4a}
  \begin{aligned}
 & \Ddot{\xi}_{q}+\omega_{q}^2
 \xi_{q}=-3\nu\omega_q E_{q_0}\cos^2{(\hat{\omega}t)}\sum\limits_{p}
 C_{q,q_0,q_0,p}\omega_p\xi_{p}\\
 &-d \omega_q\sum\limits_p\omega_p K_{p,q}\xi_{p}+O(\nu^2,\nu d, d^2)
 \end{aligned}
\end{equation}
The strongest instability is due to primary parametric resonance in
\eqref{eq4a} and involves a pair of the resonant modes $\tilde{q},\tilde{p}=q_0\pm1$. Omitting non-resonant and $O(\nu^2,\nu d, d^2)$ terms it is reduced to 
  \begin{equation}
  \label{eq4b}
  \left\{\begin{aligned}
  & \Ddot{\xi}_{\tilde{q}}+\omega_{\tilde{q}}^2(1+d K_{\tilde{q},\tilde{q}})
 \xi_{\tilde{q}}=-3\nu\omega_{\tilde{q}}\omega_{\tilde{p}}E_{q_0}\cos^2{(\hat{\omega}t)}
 \xi_{\tilde{p}}\\
 & \Ddot{\xi}_{\tilde{p}}+\omega_{\tilde{p}}^2(1+d K_{\tilde{p},\tilde{p}})
 \xi_{\tilde{p}}=-3\nu\omega_{\tilde{p}}\omega_{\tilde{q}}E_{q_0}\cos^2{(\hat{\omega}t)}
 \xi_{\tilde{q}}\\
 \end{aligned}
 \right.
\end{equation}
Thus, the disorder does not create new resonant terms, its impact being confined to the QB and resonant modes frequency shifts. The analysis analogous to \cite{we_qb} yields the instability threshold $\beta^*$, its mean and variance:
\begin{equation}
\label{eq5}
\begin{aligned}
&\beta^*=\beta_0^*\left(1-\frac{2d(N+1)^2}{\pi^2}\Delta K\right), \\
 & \left\langle \beta^*\right\rangle=\beta_0^*, \ \sigma_{\beta^*}=2\sigma_\kappa D\sqrt{N+1}/E_{q_0},
\end{aligned}
\end{equation}
where the disorder-free value is $\beta_0^*=\frac{\pi^2}{6 E_{q_0}(N+1)}$ and $\Delta K=K_{\tilde{q},\tilde{q}}-2K_{q_0,q_0}+K_{\tilde{p},\tilde{p}}$. It agrees well with the numerical results (Fig.\ref{fig2}(b)).

One may ask now, which particular realizations favor or disfavor stability? Furthermore, if some receipts are disclosed, can they be used in controlling the energy flow in the mode space? 
The disorder-determined part of (\ref{eq5}) can be rewritten as $\Delta K =-\frac{4}{\sqrt{N+1}}\sum\limits_{n=1}^{N+1}\kappa_n\cos{\frac{\pi 2q_0(n-0.5)}{N+1}}\sin^2{\frac{\pi(2n-1)}{2(N+1)}}$. It is linear with respect to $\kappa_n$, thus we can represent it as a sum of spatial Fourier components, their contributions being additive. Thus, consider $\kappa_n=0.5\cos{(\frac{\pi p (n-0.5)}{N+1}+\varphi)}$, where $\varphi$ is the phase shift. It is natural to expect the minimum of $\Delta K$ (and the maximal gain in stability), when $p=2q_0$, and it indeed yields $\Delta K=0.5\sqrt{N+1}\cos{\varphi}$, and the maximum $\beta^*=\beta_0^*\left(1+D(N+1)^2/\pi^2\right)$ for $\varphi=0$. Immediately, a high sensitivity on $\varphi$ is seen: the zero shift $\beta^*=\beta_0^*$ for $\varphi=\pm\pi/2$; the minimum  $\beta^*=\beta_0^*\left(1-D(N+1)^2/\pi^2\right)$ for $\varphi=-\pi$. The effect of $p=2q_0$ on adjacent QBs $q_0'=q_0\pm1$ is twice as small and reverse: for example, if $\varphi=0$ then $\beta^*=\beta_0^*\left(1-D(N+1)^2/(2\pi^2)\right)$. Remarkably, while for $p=2 q_0$ extremal shifts correspond to $\varphi=0, \pi$ and zero ones to $\varphi=\pm\pi/2$, for $p=2q_0\pm1$ the zero shift appears for $\varphi=0, \pi$, and the extrema for $\varphi=\pm\pi/2$: $\beta^*=\beta_0^*\left(1\mp 8 D(N+1)^2/(3\pi^3)\right)$. 

These results are illustrated in Fig.\ref{fig5}, and show a good correspondence to the numerically determined QB stability. That is, depending on the phase $\varphi$, the spatially harmonic modulation of springs elasticities with the wave number $p=2 q_0$, may significantly augment, weaken, or leave the stability intact (Fig.\ref{fig5}(a)). Modulations with $p=2q_0\pm2$ change the stability reversely and with twice a smaller amplitude for the same $\varphi$, and those with $p=2q_0\pm1$ -- just a bit weaker than $2q_0$, but with a $\pi/2$ shift in $\varphi$ (Fig.\ref{fig5}(b)). Notably, modulations with other wave numbers have only a minor effect. Therefore, the spatial Foirier components with $p\in[2q_0-2,2q_0+2]$ of $\{\kappa_n\}$ are decisive for the $q_0$-QB stability. 

These findings suggest a possibility of controlling the energy flow between modes. Indeed, by imposing a proper periodic modulation of the linear elasticity one can destabilize certain QB excitations and (i) promote equipartition or (ii) stabilize others, where the energy will be radiated; new QBs may also be subject to the same procedure to arrange the further energy flow. Experimentally, elasticity modulations could be achieved, for example, by laser heating, either as harmonic or spot impurities, like it was designed to control discrete breathers location in cantilever arrays \cite{Sievers}.

In conclusion, we have demonstrated, that the concept of QBs can be successfully applied to analyzing nonlinear vibrational modes in weakly disordered lattices. They essentially retain exponential localization and stability in the mode space, if the disorder is sufficiently small. We show, that the stability trend depends sensitively on a particular realization of disorder, and deliberately created inhomogeneities offer a promising technique of controlling the energy flow between nonlinear modes. We expect that these ideas and methods to be applicable to a variety of nonlinear weakly disordered lattices -- and we have already applied them to the DNLS chain (to be reported elsewhere) -- including the contexts of a different source of disorder (masses, nonlinearities), higher dimensions, and quantum arrays. The results on the nonlinear modes sustainability, stability, and controlling are strongly expected to be in demand from experiments and applications.

We thank S. Flach for stimulating and extremely valuable discussions.% and O. Kanakov for useful comments.

\end{document}